# An analysis of semiclassical radiation from single particle quantum currents shows surprising results

Mark P. Davidson[a]

[a]Spectel Research Corp., 807 Rorke Way, Palo Alto,  CA  94303, USA
Email: mdavid@spectelresearch.com

**ABSTRACT**. Classical electromagnetic radiation from quantum currents and densities are calculated.  For the free Schrödinger equation with no external force it's found that the classical radiation is zero to all orders of the multipole expansion.  This is true of mixed or pure states for the charged particle.  It is a non-trivial and surprising result.  A similar result is found for the Klien-Gordon currents when the wave function consists of only positive energy solutions.  For the Dirac equation it is found that radiation is suppressed at lower frequencies but is not zero at all frequencies.  Implications of these results for the interpretation of quantum mechanics are discussed.

## 1  Introduction

Semiclassical radiation theory has successfully described many phenomena in quantum optics and quantum electrodynamics where one might have thought that a description in terms of photons and standard quantum electrodynamics (QED) would have been more appropriate [1-3].  The consensus is that semiclassical methods fall far short of describing all of the phenomenon successfully calculated with renormalization methods and QED [3].  Nevertheless, there is a great deal of interest in applying and extending semiclassical methods.  It is surprising then that there does not seem to be in the literature an analysis of the radiation that would be produced by a quantum single particle current coupled to the classical electromagnetic field. The present work grew out of a desire to answer the following simple question.  Does a free electron described by Schrödinger's equation radiate if the electromagnetic field is treated classically and the charge density and currents are taken to be proportional to the probability density and currents of the Schrödinger wave? After checking the literature, it was found that apparently this question has not been addressed.

The reader might reasonably ask why is this question interesting.  After all, wouldn't one expect the free particle not to radiate?  Isn't this required for self-consistency of quantum mechanics?  Wouldn't one expect the classical correspondence principle to dictate this?  Of course these statements are true.  But it does not at all follow that treating the radiation classically should give a zero result.  Arguing against a null result is the analysis [4] which found that the electromagnetic self force on a quantum particle is not generally zero, even for a free particle.  One might expect from this result that the radiation from a free-particle Schrödinger current would be non-zero as well.  Moreover, consider the following gedanken experiment.  Imagine superimposing two wave packets. Let each wave packet be a plane wave modulated by a broad Gaussian envelope.  If the wavefunctions are broad enough, then their energy and momenta can have a small uncertainty.  Assume this to be the case, and let the mean energies of the two waves be different.  When these quasi-plane waves are superimposed they will in-terfere similar to laser beams interfering in optics. The wave function would have the form (treating the quasi-plane waves as perfect plane waves for the sake of argument)

$$\Psi(x,t) = A\exp(i\nu_A t - i\mathbf{k}_A \cdot \mathbf{x}) + B\exp(i\nu_B t - i\mathbf{k}_B \cdot \mathbf{x}) \tag{1}$$

And the density would then be

$$\rho(\mathbf{x},t) = q\Psi^*\Psi =$$
$$q|A|^2 + q|B|^2 + 2q\,\mathrm{Re}\left[A^*B\exp\left(i(\nu_B - \nu_A)t - i(\mathbf{k}_B - \mathbf{k}_A)\cdot\mathbf{x}\right)\right] \tag{2}$$

This expression for the density clearly shows an oscillatory time dependence with frequency $(\nu_B - \nu_A)$. A similar oscillation is found in the Schrödinger current. One would expect then from this sinusoidal dependence in both the density and the current that the radiation calculated would be non-zero. In fact, one can imagine adding more plane waves to the mix and getting quite complex time dependence. Looked at this way, one expects radiation even for a free particle. So which argument is correct?

It is shown here that there is no radiation whatsoever from the charge density and current of a free Schrödinger particle to all orders of the multipole expansion and for any state whatsoever provided certain regularity assumptions are satisfied. Then it is shown that if there are forces present, the radiation agrees with Larmor's formula to lowest order, but that there are also extra correction terms which depend on the wave function.

Another reason why these results are interesting relates to the desire to obtain a deeper realistic understanding of quantum mechanics. In the model of De Broglie [5] for example, the Schrödinger wave is related to a real physical wave which has a soliton-like singularity. The soliton motion is chaotic, but statistically described by the Schrödinger wave which is a normalized version of a real wave. The results obtained here suggest that electrodynamic forces will play an important role in the ultimate derivation and understanding of the double solution model of quantum mechanics. They suggest that the quantum mechanical potential makes a free particle diffuse in just such a way that there is no net radiation when when the Schrodinger probability and charge currents are treated as real currents. Surely this is not a coincidence that the radiation vanishes, as we show below, to all orders of the multipole expansion.

## 2 Radiation Analysis for a free Schrödinger particle

Consider first the free particle Schrödinger equation. It shall be assumed that the wave function in momentum representation has no support for any values of momentum which would correspond to motions faster than the speed of light c. This is not ensured by Schrödinger's equation automatically. Without this constraint the current and charge distribution for the particle would not be sufficiently localized and the non-radiation result would not follow. It shall first be demonstrated that there is no classical radiation in this case, provided liberal regularity assumptions are made about the wave function, and where the charges and current densities from Schrödinger's equation

$$-\frac{\hbar^2}{2m}\Delta\Psi = i\hbar\frac{\partial\Psi}{\partial t} \tag{3}$$

are used as sources for classical electromagnetic fields. The charge and current densities are

$$\rho(\mathbf{x},t) = q\Psi^*\Psi; \quad \mathbf{J}(\mathbf{x},t) = \frac{q\hbar}{2mi}\{\Psi^*\nabla\Psi - \Psi\nabla\Psi^*\} \tag{4}$$

$$J^\mu = (c\rho, \mathbf{J}), \quad x^\mu = (ct, \mathbf{x}) \tag{5}$$

Now consider the electromagnetic field generated by these sources.

$$\partial_\mu F^{\mu\nu} = \frac{4\pi}{c}J^\nu; \quad F^{\mu\nu} = \partial^\mu A^\nu - \partial^\nu A^\mu \tag{6}$$

$$A^\mu = (\Phi, \mathbf{A}) \tag{7}$$

(the metric is timelike, ie. $g^{00} = 1$). The fields are determined up to addition of an arbitrary free field, but this free field will not contribute to any radiation, and so it can be chosen to be zero for convenience. Working in the Lorentz gauge ($\partial^\mu A_\mu = 0$) one then has the classical result

$$A^\mu(\mathbf{x},t) = \frac{1}{c}\int \frac{J_\mu(\mathbf{x}',t-\frac{R}{c})}{R}d^3x'; \quad R = |\mathbf{x}-\mathbf{x}'| \tag{8}$$

The constraint on the momentum is

$$\rho(\mathbf{p}) = \Psi^*(\mathbf{p})\Psi(\mathbf{p}) = 0, \text{ for } \left|\frac{\mathbf{p}}{m}\right| > (1-\delta)c \text{ for some } \delta > 0 \tag{9}$$

With this constraint, the phase velocities of the waves making up the wavefunction have a maximum absolute value of

$$v_{max} = (1-\delta)c < c \tag{10}$$

It is still possible to localize the Schrödinger particle with this constraint on the phase velocity, and all the higher moments calculated below can be finite provided the wave function is infinitely differentiable as a function of **p**. The magnetic field is derived from the vector potential by

$$\mathbf{B}(\mathbf{x},t) = \nabla \times \mathbf{A} = \nabla \times \frac{1}{c}\int \frac{\mathbf{J}(\mathbf{x}',t-\frac{R}{c})}{R} d^3x' \tag{11}$$

Define a unit vector

$$\hat{\mathbf{n}} = \frac{\mathbf{x}-\mathbf{x}'}{|\mathbf{x}-\mathbf{x}'|} \tag{12}$$

The curl can now be evaluated as follows

$$\mathbf{B}(\mathbf{x},t) = \frac{1}{c}\int \hat{\mathbf{n}} \times \frac{\partial}{\partial R}\frac{\mathbf{J}(\mathbf{x}',t-\frac{R}{c})}{R} d^3x' \tag{13}$$

$$\mathbf{B}(\mathbf{x},t) = \frac{1}{c}\int \hat{\mathbf{n}} \times \left[-\frac{1}{R^2}\mathbf{J}(\mathbf{x}',t-R/c) - \frac{1}{Rc}\dot{\mathbf{J}}(\mathbf{x}',t-R/c)\right] d^3x' \tag{14}$$

In evaluating the radiation emitted, the limit where $|\mathbf{x}| \to \infty$ is taken, and therefore the leading behavior of **B** is all that need be kept.

$$R = \sqrt{\mathbf{x}^2 + \mathbf{x}'^2 - 2\mathbf{x}\cdot\mathbf{x}'} \tag{15}$$

Define

$$R_0 = |\mathbf{x}| \tag{16}$$

And so to leading order in $R_0$

$$\hat{n} = \frac{\mathbf{x}}{|\mathbf{x}|} \tag{17}$$

$$R = R_0 - \hat{\mathbf{n}} \cdot \mathbf{x}' \tag{18}$$

And so to leading order in $R_0$

$$\mathbf{B}(\mathbf{x},t) = -\hat{\mathbf{n}} \times \frac{1}{c^2 R_0} \int \dot{\mathbf{J}}(\mathbf{x}',t - \frac{R_0 - \hat{\mathbf{n}} \cdot \mathbf{x}'}{c}) d^3x' \tag{19}$$

Now expand in a Taylor series. It is assumed that **J** is infinitely differentiable as a function of time and that the Taylor series converges

$$\mathbf{B}(\mathbf{x},t) = -\hat{\mathbf{n}} \times \frac{1}{c^2 R_0} \sum_{m=1}^{\infty} \int \left( \frac{\partial^m}{\partial t^m} \mathbf{J}(\mathbf{x}',t - \frac{R_0}{c}) \right) \left( \frac{\hat{\mathbf{n}} \cdot \mathbf{x}'}{c} \right)^{m-1} /(m-1)! \, d^3x' \tag{20}$$

This is essentially the multipole expansion for the field. Now insert the Schrödinger current (4) into this equation, and assume that the order of summation and integration can be interchanged. One must evaluate the following integrals

$$\mathbf{I}_m(t_0) = \int \mathbf{J}(\mathbf{x}',t_0) \left( \frac{\hat{\mathbf{n}} \cdot \mathbf{x}'}{c} \right)^{m-1} d^3x', \text{ where } t_0 = t - R_0/c \tag{21}$$

In terms of which **B** may be written to leading order in $R_0$

$$\mathbf{B}(x,t) = -\hat{\mathbf{n}} \times \frac{1}{c^2 R_0} \sum_{m=1}^{\infty} \frac{\partial^m}{\partial t_0^m} \mathbf{I}_m(t_0) \left( \frac{1}{c} \right)^{m-1} /(m-1)! \tag{22}$$

$\mathbf{I}_m$ takes the form

$$\mathbf{I}_m(t_0) =$$
$$\frac{q\hbar}{2mi} \int \left\{ \Psi^*(\mathbf{x}',t_0)\nabla\Psi(\mathbf{x}',t_0) - \Psi(\mathbf{x}',t_0)\nabla\Psi^*(\mathbf{x}',t_0) \right\} \left( \hat{\mathbf{n}} \cdot \mathbf{x}' \right)^{m-1} d^3x' \tag{23}$$

this may be written as

$$\mathbf{I}_m = \frac{q}{2m} \int \Psi^*(\mathbf{x}',t_0) \left\{ \mathbf{P}(\hat{\mathbf{n}} \cdot \mathbf{x}')^{m-1} + (\hat{\mathbf{n}} \cdot \mathbf{x}')^{m-1} \mathbf{P} \right\} \Psi(\mathbf{x}',t_0) d^3x' \tag{24}$$

where

$$\mathbf{P} = -i\hbar\nabla \tag{25}$$

Now comes a marvelous surprise. Transform (24) by switching to the Heisenberg representation. The time-dependent Heisenberg operators are

$$\mathbf{P}(t) = \mathbf{P}, \quad \mathbf{x}(t) = \mathbf{x} + t\mathbf{P}/m \tag{26}$$

And the quantum expectations $\mathbf{I_m}$ are then

$$\mathbf{I}_m(t_0) = \frac{q}{2m}\int \Psi^*(\mathbf{x},0)\left\{\mathbf{P}(t_0)\left(\hat{\mathbf{n}}\cdot\mathbf{x}'(t_0)\right)^{m-1} + \left(\hat{\mathbf{n}}\cdot\mathbf{x}'(t_0)\right)^{m-1}\mathbf{P}(t_0)\right\}\Psi(\mathbf{x},0)d^3x' \tag{27}$$

$$\mathbf{I}_m(t_0) = \text{Polynomial of order (m-1) in the variable } t_0 \tag{28}$$

Therefore it follows that

$$\frac{\partial^m}{\partial t^m}\mathbf{I}_m(t_0) = 0 \tag{29}$$

And hence to leading order in $R_0$

$$\mathbf{B}(\mathbf{x},t) = 0 \tag{30}$$

More precisely one has that

$$\lim_{|\mathbf{x}|\to\infty}|\mathbf{x}|^{2-\varepsilon}\mathbf{B}(\mathbf{x},t) = 0, \text{ for all } \varepsilon > 0 \tag{31}$$

The electric field in a region which is far from any charges or currents can be calculated by

$$\mathbf{E} = -\hat{\mathbf{n}}\times\mathbf{B} \tag{32}$$

And therefore the integrated Poynting vector over a spherical surface S of radius $R_0$ vanishes in the limit

$$\lim_{R_0\to\infty}\frac{4\pi}{c}\iint_S \mathbf{E}\times\mathbf{B}\cdot d\mathbf{S} = 0 \tag{33}$$

And so it can be concluded that the classical radiation calculated for the free Schrödinger particle is zero.

The free particle wave function will spread out as time passes and eventually would reach any finite radius in extent. Thus, if one were to wait long enough, there there would eventually be electromagnetic energy flowing through any sphere whose radius is finite but fixed in time. But this is clearly not electromagnetic radiation, because it is propogating at the group velocity of the Schrödinger wave which is less than c by assumption. The electromagnetic radiation must propogate at c, and therefore one can choose the integration sphere to be large enough so that the radiation has had time to reach it but the spreading wave function has not had sufficient time.

## 3  Mixed States also gives zero radiation in semiclassical approximation

The preceding analysis showed that both E and B fall off more quickly than 1/R for all orders of the multipole expansion. If we were to simply add several such currents together, as in a mixed state, then E and B would still fall off more quickly than 1/R and so the result of no radiation in this semiclassical calculation would still hold for mixed states like this. As a mathematical curiosity, it follows that adding mixed states with different values of $\hbar$ together would also give a zero result. It is an interesting question as to whether this exhausts all non-trivial time varying currents that have the property that they don't radiate.

## 4  Radiation Analysis for an accelerating Schrödinger particle - Larmor's formula and higher order terms

Consider the first term in the expansion for **B** (22)

$$\mathbf{I}_1 = \frac{q}{m}\mathbf{P}(t_0) \tag{34}$$

$$\mathbf{B}(x,t) \cong -\hat{\mathbf{n}} \times \frac{1}{c^2 R_0}\frac{q}{m}\int \Psi^*(\mathbf{x}')\dot{\mathbf{P}}(t-R_0/c)\Psi(\mathbf{x}')d^3x' \tag{35}$$

And again using (32) one has

$$S = \frac{c}{4\pi}\mathbf{E}\times\mathbf{B} = \frac{q^2}{4\pi c^3 R_0^2}\mathbf{a}^2\sin^2(\theta)\hat{\mathbf{n}} \tag{36}$$

Where

$$\mathbf{a} = \frac{1}{m}\int \Psi^*(\mathbf{x}')\dot{\mathbf{P}}(t-R_0/c)\Psi(\mathbf{x}')d^3x' \tag{37}$$

And the total power radiated is

$$P_R = \int_0^\pi \frac{q^2}{4\pi c^3 R_0^2} \mathbf{a}^2 \sin^2(\theta) R_0^2 \, 2\pi \sin(\theta) d\theta \tag{38}$$

Which simplifies to

$$P_R = \frac{2}{3} \frac{q^2}{c^3} \mathbf{a}^2 \tag{39}$$

This is just Larmor's formula. Keeping the higher order terms in the expansion for **B** (22) yields correction terms to the Larmor formula. These corrections depend on the specific functional form of the wave function through the $I_m$. But even the lowest order term already differs from QED [6].

## 5  Radiation analysis for a Newtonian ensemble

Next consider an ensemble of Newtonian particles which are constrained to move with constant velocity. Such a system can be described by a phase space distribution function $\rho(\mathbf{x}, \mathbf{v}, t)$ which satisfies the Liouville equation

$$\frac{\partial \rho(\mathbf{x}, \mathbf{v}, t)}{\partial t} + \mathbf{v} \cdot \nabla \rho(\mathbf{x}, \mathbf{v}, t) = 0 \tag{40}$$

Since by assumption the velocities of the particles don't change with time. The current density and charge density in this case are given by

$$\rho(\mathbf{x}, t) = \int \rho(\mathbf{x}, \mathbf{v}, t) d^3 \mathbf{v}; \quad \mathbf{j}(\mathbf{x}, t) = \int \mathbf{v} \rho(\mathbf{x}, \mathbf{v}, t) d^3 \mathbf{v} \tag{41}$$

The $I_m$ may be calculated by making the substitution

$$\mathbf{x}(t) = \mathbf{x}(0) + \mathbf{v} t \tag{42}$$

into the integrals for $I_m$. Again, as for the free particle Schrödinger case, the $I_m$ are polynomials of order (m-1) in the time variable and thus the radiation will again be zero. This is not a particularly surprising result however since there is no interference in this Newtonian case, and it is the interference terms that give rise to oscillations in the Schrödinger case. In both the Newtonian and the Schrödinger case, we see that it is the fact that position enters as a first order function of time that leads to the zero radiation result. In the Schrödinger case the coordinate **x(t)** is an operator which is linear in time whereas in the Newtonian case it is simply a classical vector.

A more realistic problem would be a Newtonian ensemble in which the particles are allowed to accelerate under their mutual interactions. Such a system would undoubtedly radiate.

## 6  Semiclassical radiation analysis for the Klein-Gordon equation

Consider the free Klein-Gordon equation

$$\left[\partial^\mu\partial_\mu + \frac{m^2c^2}{\hbar^2}\right]\Phi = 0 \tag{43}$$

The four current is given by

$$J_\mu = \frac{q}{2mi}\left[\Phi^*\partial_\mu\Phi - \Phi\partial_\mu\Phi^*\right] \tag{44}$$

The constraint (9) need not be imposed on the Klein-Gordon case since the velocities are automatically subluminal. The wave function is in general a superposition of positive and negative energy terms

$$\Phi = \Phi_+ + \Phi_- \tag{45}$$

The time evolution for these two components is

$$\begin{aligned}\Phi_+(\mathbf{x},t_0) &= \exp(-ic\sqrt{\mathbf{p}^2 + m^2c^2}\,t_0/\hbar)\Phi_+(\mathbf{x},0) \\ \Phi_-(\mathbf{x},t_0) &= \exp(+ic\sqrt{\mathbf{p}^2 + m^2c^2}\,t_0/\hbar)\Phi_-(\mathbf{x},0)\end{aligned} \tag{46}$$

Where p is again given by (25).

If the wavefunction has only a positive energy term, then in this case one can write a Heisenberg time dependent position operator

$$\mathbf{x}_+(t_0) = \exp(+ic\sqrt{\mathbf{p}^2 + m^2c^2}\,t_0/\hbar)\mathbf{x}\exp(-ic\sqrt{\mathbf{p}^2 + m^2c^2}\,t_0/\hbar) \tag{47}$$

Which simplifies to

$$\mathbf{x}_+(t_0) = \mathbf{x} + \frac{\mathbf{p}}{\sqrt{\frac{\mathbf{p}^2}{c^2} + m^2}}t_0 \tag{48}$$

And once again, because $x_+(t_0)$ is linear in $t_0$, the $I_m$ will be polynomials of order (m-1) in $t_0$, so that if the wavefunction contains only positive energy terms, there will again be no radiation. Likewise if the wavefunction contains only negative energy terms the classical radiation will vanish.

When the wave function contains both positive and negative energy terms, it appears that there will be classical radiation, or at least the previous arguments which lead to zero radiation are no longer valid and there does not seem to be another mechanism to suppress the radiation. Even for mixed wavefunctions however, the radiation at lower frequencies will be suppressed if one considers time averaged currents over a time long compared to $\hbar/(4\pi mc^2)$. In this case the cross terms involving + and − wavefunctions in the expression for the current density will average out to be nearly zero.

## 7 The Dirac Equation

The free-particle Dirac equation may be written (using the notation of [7]chapter 1.4).

$$i\hbar \frac{\partial \Psi}{\partial t} = \left(c\boldsymbol{\alpha} \cdot \mathbf{p} + \beta mc^2\right)\Psi \tag{49}$$

The Hamiltonian is given by the operator

$$H = \left(c\boldsymbol{\alpha} \cdot \mathbf{p} + \beta mc^2\right) \tag{50}$$

And the charge and current densities are

$$\rho = q\sum_{\sigma=1}^{4} \Psi_\sigma^* \Psi_\sigma; \quad j^k = cq\sum_{\sigma=1}^{4} \Psi_\sigma^* \alpha^k \Psi_\sigma \tag{51}$$

The time dependent form for the coordinate **x**(t) shows oscillatory motion (zitterbewegung). Both H and **p** are independent of time though, and the coordinate satisfies the equation

$$\mathbf{x}(t) = \mathbf{x}(0) + c^2 \mathbf{p} H^{-1} t + \frac{\hbar^2}{4}\ddot{\mathbf{x}}(0) H^{-2}(1 - e^{-i2Ht/\hbar}) \tag{52}$$

Where

$$\ddot{\mathbf{x}}(0) = \frac{2c}{i\hbar}(\boldsymbol{\alpha} H - c\mathbf{p}); \quad \dot{\mathbf{x}}(0) = c\boldsymbol{\alpha} \tag{53}$$

Equation (22) can be used to calculate B in this case provided the current density for the Dirac equation is used. The expression for $I_m$ becomes

$$I_m(t_0) = cq\int \sum_{\sigma=1}^{4} \Psi_\sigma^*(\mathbf{x}',t_0) c\boldsymbol{\alpha}\left(\hat{\mathbf{n}} \cdot \mathbf{x}'\right)^{m-1} \Psi_\sigma(\mathbf{x}',t_0) d^3x' \tag{54}$$

The operators **α** and **x** commute, and thus in the Heisenberg representation where they become time dependent, they must also commute at the same time. Therefore, one may shift the time dependence to the operators without regard to their order and obtain

$$I_m(t_0) = cq \int \sum_{\sigma=1}^{4} \Psi_\sigma^*(\mathbf{x}',0) \dot{\mathbf{x}}(t_0) \left( \hat{\mathbf{n}} \cdot \mathbf{x}'(t_0) \right)^{m-1} \Psi_\sigma(x',0) d^3 x' \qquad (55)$$

It is clear from this expression that all of the $I_m$ have exponential behavior in time, and therefore the previous argument cannot be used to argue that the Dirac equation will not radiate classically. The oscillatory zitterbewegung term has an angular frequency of at least $4\pi mc^2/\hbar$, the same as the oscillation between the cross terms for the Klein-Gordon equation, and so one expects some radiation here at this and higher frequencies. For much lower frequencies the oscillatory term can be safely ignored and the classical radiation will be greatly suppressed.

## 8  Conclusion

It has been shown that the charge density and currents for a free particle Schrödinger equation do not radiate when taken as the source for a classical electromagnetic field, and when liberal regularity assumptions are made about the differentiability and localizability of the densities. Larmor's formula is also derived as a first approximation to an infinite series when the Schrödinger particle has a non-vanishing acceleration. The same is true for mixed states of free particles.

It was shown that an ensemble of Newtonian charged particles moving with constant velocity also do not radiate as expected.

For the free Klein-Gordon equation it was shown that there is no classical radiation when the wavefunction is purely positive energy or purely negative energy, and that even for mixed wavefunctions the radiation is suppressed at lower frequencies.

For the Dirac equation it was shown how the radiation is suppressed at lower frequencies, but that the zitterbegung seems to lead to radiation at high frequencies. Perhaps this could be cancelled by a spin contribution that was not included in the analysis.

The schrödinger equation and to a lesser extent the relativistic wave equations seem to belong to a very special class of fluid-like models for which, though the charged flow is quite time dependent, the classical radiation is nevertheless zero for the free particle.

These results are consistent with the classical correspondence principal, and so they are reassuring from the point of view of the logical consistency of quantum mechanics. One might argue that the two cases found where the free particles do seem to radiate, the Klein-Gordon equation with mixed states and the Dirac equation both at high frequencies, are simply examples where the semiclassical radiation picture breaks down.

It is tempting to argue that the results of this paper add support to the hypothesis that the origins of quantum behavior are somehow deeply connectd to classical electromagnetic theory and to the self interaction of charged particles as in [8-11]. As has been shown here, the extremely complex time dependence of the densities and currents in quantum mechanics are just such as to provide for complete cancellation of all radiation no matter what the wave function is for a free particle.